\documentclass{jfm}

\usepackage{amssymb,bm,dsfont,mathtools,xcolor}
\usepackage{graphicx,scrextend}
\usepackage{enumerate}

\DeclarePairedDelimiter{\set}{\{}{\}}

\DeclareMathOperator{\sign}{sgn}

\newcommand{\eps}{\varepsilon}
\newcommand{\grad}{\nabla}
\renewcommand{\div}{\nabla\cdot}

\renewcommand{\a}{\bm{a}}

\newcommand{\n}{\bm{n}}
\renewcommand{\p}{\bm{p}}

\renewcommand{\u}{\bm{u}}

\newcommand{\x}{\bm{x}}

\newcommand{\A}{\bm{A}}
\newcommand{\B}{\bm{B}}

\newcommand{\D}{\bm{D}}
\newcommand{\E}{\bm{E}}

\newcommand{\I}{\bm{I}}

\newcommand{\W}{\bm{W}}

\newcommand{\bSigma}{\bm\Sigma}

\newcommand{\cE}{\mathcal{E}}
\newcommand{\cF}{\mathcal{F}}

\newcommand{\cH}{\mathcal{H}}
\newcommand{\cI}{\mathcal{I}}

\newcommand{\cK}{\mathcal{K}}

\newcommand{\ce}{{\rm ce}}
\DeclareMathOperator*{\argmin}{argmin}

\graphicspath{{figures/}}

\begin{document}

\title{Variational bounds and nonlinear stability of an active nematic suspension}
\author{Scott Weady\aff{1}
\corresp{\email{sweady@flatironinstitute.org}}}
\affiliation{\aff{1}Center for Computational Biology, Flatiron Institute, New York, NY, 10010, USA}
\date{\today}

\maketitle

\begin{abstract}
We use the entropy method to analyze the nonlinear dynamics and stability of a continuum kinetic model of an active nematic suspension. From the time evolution of the relative entropy -- an energy-like quantity in the kinetic model -- we derive a variational bound on relative entropy fluctuations that can be expressed in terms of orientational order parameters. From this bound we show isotropic suspensions are nonlinearly stable for sufficiently low activity, and derive upper bounds on spatiotemporal averages in the unstable regime that are consistent with fully nonlinear simulations. This work highlights the self-organizing role of activity in particle suspensions, and places limits on how organized such systems can be.
\end{abstract}

\section{Introduction}

Suspensions of active particles, such as swimming microorganisms or microtubules mixed with molecular motors, are a canonical class of active matter. When the number of suspended particles is large, these active suspensions can transition into large scale collective motion characterized by persistent unsteady flows \citep{Sanchez2012,Wensink2012,Dunkel2013}, concentration fluctuations \citep{Narayan2007,Liu2021}, and long-range correlations \citep{Dombrowski2004,Peng2021}. Owing to their visual similarities with inertial turbulence, these dynamics are often called active or bacterial turbulence \citep{Alert2022}, and many methods from classical turbulence theory have been used to understand their structure.

Continuum models based on partial differential equations are powerful tools for studying active suspensions. One popular model is the Doi-Saintillan-Shelley (DSS) kinetic theory \citep{Saintillan2008}, which describes the configuration of particle positions and orientations through a continuous distribution function. The DSS kinetic theory is similar to well-studied models of passive polymer suspensions such as the Doi theory \citep{Doi1986}, however it is distinguished by particle motility and a consequent ``active'' stress. Motility and the corresponding stress couple the translational and orientational degrees of freedom in a novel way that can cause instabilities and drive large concentration fluctuations. 

Various works on the DSS theory and related models have explored the impact of activity on mixing \citep{Albritton2023,Zelati2023}, correlations \citep{Stenhammar2017,Skultety2020}, and stability \citep{Simha2002,Hohenegger2010,Ohm2022}, primarily at the linear or weakly nonlinear levels or below the transition to instability. At the fully nonlinear level, however, different tools are necessary. For classical problems in fluid mechanics, such as boundary driven or natural convective flows, the energy method is a powerful approach that can be used to prove existence, uniquenes, and/or stability of a broad class of solutions \citep{Straughan1992,Majda2001}. In the context of stability, one typically looks at the time evolution of the perturbation kinetic energy $\cK(t) = \int_\Omega |\delta\u|^2/2 ~ d\x$, where $\delta\u$ is the velocity deviation from a steady state, of arbitrary magnitude, and $\Omega$ is the domain of interest \citep{Doering1995}. Other non-negative integral quantities may also be used where the method is sometimes called the Lyapunov method or the entropy method depending on the chosen functional. Entropy methods in particular are common in the analysis of Fokker-Planck-type equations, such as the Doi-Saintillan-Shelley theory, owing to their probabilistic description \citep{Arnold2004,Chen2013}. Tools related to the energy method, such as the background method, can also provide bounds on time-averaged quantities for turbulent flows \citep{Doering1992,Fantuzzi2022}. Such bounds provide insight into turbulence and hydrodynamic stability at the fully nonlinear level.

Drawing on analogies between active and inertial turbulence, it is natural to ask under what conditions active suspensions are stable and to quantify how unsteady they may be. In this paper, we address these questions in the context of the Doi-Saintillan-Shelley kinetic theory for a suspension of immotile, yet active, particles --- an example of an active nematic suspension \citep{Gao2017,Doostmohammadi2018}. For ease of discussion we focus on two-dimensional suspensions, however the results extend to three dimensions with few modifications. The key quantity of interest here is the relative configuration entropy, which plays the role of an energy in this model. The paper is outlined as follows. We first describe the DSS kinetic theory for a dilute suspension of active particles. We then derive a variational bound on relative entropy fluctuations which is local in space and can be analyzed using elementary methods. Using this bound, we derive explicit uniform-in-time bounds on the relative entropy. We then prove a sharp condition for nonlinear stability and derive bounds on time averages of orientational order parameters which hold in the turbulent regime. These results are validated against fully nonlinear simulations of the kinetic theory.

\section{The Doi-Saintillan-Shelley kinetic theory}

In this section we summarize the Doi-Saintillan-Shelley kinetic theory for an immotile active suspension, further details can be found in various references \citep{Saintillan2008,Saintillan2013}. Consider a suspension of $N$ particles in a domain $\Omega \subseteq \mathds{R}^2$, either bounded or periodic. Let $\Psi(\x,\p,t)$ describe the probability of finding a particle at position $\x \in \Omega$ with orientation $\p \in S = \set{\p\in\mathds{R}^2:|\p|=1}$ such that $\int_\Omega \int_S \Psi ~ d\p d\x = N$. Assuming the total number of particles is conserved, this distribution function satisfies a Smoluchowski equation,
\begin{gather}
    \frac{\partial\Psi}{\partial t} + \grad_x\cdot(\dot\x\Psi) + \grad_p\cdot(\dot\p\Psi) = 0,\quad\x\in\Omega,\label{eq:dPsi/dt}\\\quad \dot\x\cdot\hat\n = 0,\quad\x\in\partial\Omega,\label{eq:xdotn}
\end{gather}
where $\grad_x = \partial_{\x}$ is the spatial gradient operator, $\grad_p = (\I - \p\p)\cdot\partial_{\p}$ is the gradient operator on the unit sphere, and $\hat\n$ is the unit normal vector to the boundary. The configuration fluxes $\dot\x(\x,\p,t)$ and $\dot\p(\x,\p,t)$ describe the dynamics of a single particle and, in the dilute limit, are given by
\begin{align}
    \dot\x &= \u - D_T\grad_x\log\Psi,\label{eq:xdot}\\
    \dot\p &= (\I - \p\p)\cdot(\gamma\E + \W)\cdot\p - D_R\grad_p\log\Psi,\label{eq:pdot}
\end{align}
where $\E = (\grad\u+\grad\u^T)/2$ is the symmetric rate of strain and $\W = (\grad\u-\grad\u^T)/2$ is the vorticity tensor under the convention $(\grad\u)_{ij} = \partial u_i/\partial x_j$. (When acting on a function of space alone, we use the shorthand notation $\grad := \grad_x$.) The first equation says particles are advected by the local fluid velocity $\u(\x,t)$ and diffuse in space with diffusion coefficient $D_T$, assumed to be isotropic. The second equation describes particle rotation by velocity gradients according to Jeffery's equation, where $\gamma\in[-1,1]$ is a dimensionless geometric factor that satisfies $\gamma = -1$ for plates, $\gamma = 0$ for spheres, and $\gamma = 1$ for rods. The orientation dynamics also includes diffusion with coefficient $D_R$, again assumed to be isotropic. 

For a passive suspension, the velocity field is often prescribed and the Smoluchowski equation can be solved accordingly. However, particles, either active or passive, will induce a stress $\bSigma_a$ on the fluid that will modify the flow field. At leading order, this stress takes the form of a dipole, $\bSigma_a = \sigma_a\D$, where $\D = \langle\p\p - \I/2\rangle$ is the trace-free second moment of the distribution function with the notation $\langle f(\p) \rangle = \int_S f(\p) \Psi d\p$. The sign of $\sigma_a$ depends on the microstructure, and is positive for contractile particles and negative for extensile particles. Owing to the small scales under consideration, this stress balances the incompressible Stokes equations
\begin{gather}
    -\nu\Delta\u + \grad \Pi = \grad\cdot\bSigma_a,\quad\x\in\Omega,\label{eq:delu}\\
    \grad\cdot\u = 0,\quad\x\in\Omega,\label{eq:divu}\\
    \u = 0,\quad\x\in\partial\Omega,\label{eq:udotn}
\end{gather}
where $\Pi(\x,t)$ is the pressure which enforces the incompressibility condition (\ref{eq:divu}) and $\nu$ is the viscosity.  Note that the no-slip boundary condition $\u = 0$ implies the distribution function satisfies a homogeneous Neumann condition $\partial\Psi/\partial n = 0$ for $\x\in\partial\Omega$. We refer to Eqs. (\ref{eq:dPsi/dt})-(\ref{eq:udotn}) as the Doi-Saintillan-Shelley (DSS) kinetic theory.

Moments of the distribution function correspond to orientational order parameters which provide a useful characterization of the macroscopic dynamics. Two key parameters here are the concentration $c = \langle 1 \rangle$ and the scalar nematic order parameter $\mu$, which is the largest eigenvalue of the normalized second-moment tensor $\D/c$. Note that, because the suspension is nematic, the polarity vector $\langle\p\rangle/c$ does not appear in the dynamics.

\subsection{Non-dimensionalization}

Similar to the procedure of \citep{Ohm2022}, the distribution function is normalized by the number density $n = N/|\Omega|$ so that $\int_\Omega\int_S \Psi ~ d\p d\x = |\Omega|$, and we non-dimensionalize by a characteristic length scale $\ell_c = 1/\lambda^{1/2}$, where $\lambda$ is the principal Dirichlet eigenvalue of the Laplacian on $\Omega$. For example,  $\lambda = (2\pi/L)^2$ for a periodic box of length $L$. We also non-dimensionalize by the active time scale $t_c = \nu/n|\sigma_a|$. This yields, in addition to the geometric factor $\gamma$, two dimensionless parameters, $D_T' = (\nu\lambda/n|\sigma_a|)D_T$ and $D_R' = (\nu/n|\sigma_a|)D_R$, which are the dimensionless translational and rotational diffusion coefficients, respectively. Ignoring primes on dimensionless variables, the Smoluchowski equation (\ref{eq:dPsi/dt}) and configuration fluxes (\ref{eq:xdot}) and (\ref{eq:pdot}) keep the same form, with the dimensionless Stokes equations
\begin{gather}
    -\Delta\u + \grad \Pi = \sign(\sigma_a)\grad\cdot\D,\\
    \div\u = 0.
\end{gather}

\subsection{The relative configuration entropy}

A natural quantity that arises from the kinetic theory's probabilistic description is the relative configuration entropy,
\begin{equation}
    \cH[\Psi](t) = \int_\Omega \int_S \Psi\log\Big(\frac{\Psi}{\Psi_0}\Big) ~ d\p d\x,\label{eq:H}
\end{equation}
where $\Psi_0 = 1/|S| = 1/2\pi$ is the isotropic distribution function. This is a non-negative quantity that is zero only in the globally isotropic state, $\Psi(\x,\p) = \Psi_0$, and increases with particle alignment. Differentiating in time and using the Smoluchowski equation, one can show $\cH$ satisfies, in two dimensions,
\begin{equation}
    \frac{d\cH}{dt} = -4\gamma\sign(\sigma_a)\int_\Omega |\E|^2 ~ d\x - 4\int_\Omega \int_S D_T |\grad_x \Psi^{1/2}|^2 + D_R |\grad_p \Psi^{1/2}|^2 ~ d\p d\x,\label{eq:dH/dt}
\end{equation}
where $|\A| = \sqrt{\A:\A}$ is the Frobenius norm for matrix quantities and $|\a| = \sqrt{\a\cdot\a}$ is the Euclidean norm for vector quantities. (Note that $(\grad_x\Psi^{1/2})_i = \partial_{x_i}(\Psi^{1/2})$, and similarly for orientational gradients.) Equation (\ref{eq:dH/dt}) balances the rate of work with spatial and rotational dissipation. For suspensions with $\gamma\sign(\sigma_a) > 0$, such as contractile rods or extensile plates, the right hand side is strictly negative so that the relative entropy always decreases. On the other hand, when $\gamma\sign(\sigma_a) < 0$, as is the case for extensile rods or contractile plates, the right hand side may be positive or negative. The case $\gamma\sign(\sigma_a) > 0$ is nearly identical to models of passive suspensions, such as the classical Doi theory, and is well studied \citep{Doi1986,Constantin2005,Constantin2008}. In the remainder of this work we therefore only consider extensile suspensions in the rod limit, $\gamma\sign(\sigma_a) = -1$, for which the model exhibits complex spatiotemporal dynamics, though all of the arguments can be followed identically for any value of $\gamma$.

It is natural to ask if the relative entropy is essential here rather than another non-negative functional. To motivate this, consider for example the typical $L^2$ quantity
\begin{equation}
    \cE(t) = \frac{1}{2}\int_\Omega \int_S |\Psi - \Psi_0|^2 ~ d\p d\x.
\end{equation}
Like the relative entropy, $\cE$ is non-negative and zero only in the globally isotropic state. Differentiating in time and using the Smoluchowski equation, we arrive at
\begin{equation}
    \frac{d\cE}{dt} = \gamma\int_\Omega \int_S \E:(\p\p-\I/2) \Psi^2~ d\p d\x -\int_\Omega \int_S D_T|\grad_x\Psi|^2 + D_R|\grad_p\Psi|^2 ~ d\p d\x. 
\end{equation}
Fluctuations in $\cE$ similarly balance flow alignment with spatial and rotational dissipation, but the first term, having a quadratic dependence on $\Psi$, is significantly more challenging to estimate than the term $\int_\Omega |\E|^2 ~ d\x$ that arises in the evolution of $\cH$, even for the stable case $\gamma\sign(\sigma_a) > 0$.

\section{Bounds on fluctuations}

The relative entropy equation (\ref{eq:dH/dt}) holds for all solutions of the kinetic theory. However, through the Stokes equations, the right hand side involves terms that are nonlocally coupled in both the spatial and orientational degrees of freedom. In this section, we derive an upper bound on the fluctuation rate $d\cH/dt$ that depends only on local orientational order parameters. A key ingredient in this analysis is a variational bound on rotational dissipation that can be expressed in terms of moments of the distribution function alone. We treat each of the terms in Eq. (\ref{eq:dH/dt}) individually.

\subsection{Rate of work}

In conservative form, the Stokes equations are
\begin{gather}
    \grad\cdot(-2\E + \Pi\I) = -\grad\cdot\D,\label{eq:Stokes_div}\\
    \grad\cdot\u = 0.
\end{gather}
Dotting (\ref{eq:Stokes_div}) with $\u$ and integrating by parts gives
\begin{equation}
    2\int_\Omega \E:\E ~ d\x = \int_\Omega \E:\D ~ d\x,
\end{equation}
where we've used the fact that $\A:\B = [(\A + \A^T):\B]/2$ for any symmetric matrix $\B$. The Cauchy-Schwarz inequality then implies
\begin{equation}
    2\Big(\int_\Omega |\E|^2 ~ d\x\Big) \leq \Big(\int_\Omega |\E|^2 ~ d\x\Big)^{1/2} \Big(\int_\Omega |\D|^2 ~ d\x\Big)^{1/2},
    \end{equation}
or
\begin{equation}
\int_\Omega |\E|^2 ~ d\x \leq \frac{1}{4} \int_\Omega |\D|^2 ~ d\x.\label{eq:strain-bound}
\end{equation}
This inequality is formally sharp for constant $\D = 2\E$, which corresponds to a strain flow, though this solution is not valid on bounded domains. The advantage of the elementary inequality (\ref{eq:strain-bound}) is that it does not require the nonlocal coupling between the stress $\D$ and the fluid velocity $\u$.

\subsection{Spatial dissipation}

Using the definition $c = \int_S \Psi ~ d\p$, we have
\begin{equation}
    \begin{aligned}
    |\grad c| = \Big|\int_S \grad_x\Psi ~ d\p\Big|
    & \leq
    \int_S |\grad_x\Psi| ~ d\p
    \\ & =
    \int_S \frac{\Psi^{1/2}}{\Psi^{1/2}} |\grad_x\Psi| ~ d\p
    \\ & \leq
    \Big( \int_S \Psi ~ d\p\Big)^{1/2}\Big( \int_S \frac{|\grad_x\Psi|^2}{\Psi} ~ d\p\Big)^{1/2}
    \\ & = 
    c^{1/2} \Big(4\int_S |\grad_x\Psi^{1/2}|^2 ~ d\p\Big)^{1/2},
    \end{aligned}
\end{equation}
where we used the Cauchy-Schwarz inequality in the second to last line. Squaring both sides, dividing by $c$, and integrating over $\Omega$, we find
\begin{equation}
    \int_\Omega |\grad c^{1/2}|^2 ~ d\x \leq \int_\Omega\int_S |\grad_x\Psi^{1/2}|^2 ~ d\p d\x.\label{eq:D_T_bound}
\end{equation}
This inequality is sharp, which is shown by considering orientationally isotropic distributions of the form $\Psi = c\Psi_0$.

\subsection{Rotational dissipation}\label{sec:rotation}

Combining inequalities (\ref{eq:strain-bound}) and (\ref{eq:D_T_bound}), we have
\begin{equation}
    \frac{d\cH}{dt} \leq \int_\Omega |\D|^2 ~ d\x - 4D_T\int_\Omega |\grad c^{1/2}|^2 ~ d\x - 4D_R\int_\Omega \int_S |\grad_p\Psi^{1/2}|^2 ~ d\p d\x.\label{eq:dH/dt-rotational-dissipation}
\end{equation}
The last term, corresponding to rotational dissipation, frequently arises in entropy methods applied to other kinetic equations \citep{Arnold2004}. In these contexts there are a useful class of inequalities known as logarithmic Sobolev inequalities \citep{Gross1975}, which are of the form
\begin{equation}
\int_S f \log\Big(|S|\frac{f}{\int_S f ~ d\p}\Big) ~ d\p \leq C\int_S |\grad_p f^{1/2}|^2 ~ d\p,\label{eq:log-Sobolev}
\end{equation}
where $C$ is the log-Sobolev constant and $f \in H^1(S)$ with $f > 0$. For arbitrary distributions in two dimensions, the optimal constant is $C = 2$, though this constant can be improved under assumptions on the distribution function (see, for example, \cite{Dolbeault2016,Brigati2022}). Considering the case $f = \Psi$ and assuming $\Psi \in H^1(S)$, inequality (\ref{eq:log-Sobolev}) implies
\begin{equation}
    \cH -\int_\Omega c\log c ~ d\x = \int_\Omega \int_S \Psi \log\Big(\frac{\Psi}{c\Psi_0}\Big) ~ d\p d\x \leq C \int_\Omega \int_S |\grad_p \Psi^{1/2}|^2 ~ d\p d\x.\label{eq:log-Sobolev-Psi}
\end{equation}
While this inequality holds for $C = 2$ in general, we can sharpen the constant using constraints on moments of $\Psi$. 

Now let $\p = (\cos\theta,\sin\theta)^T$ where $\theta\in[0,2\pi)$ is the polar angle. Because $c$ and $\D$ appear in the upper bound (\ref{eq:dH/dt-rotational-dissipation}), the distribution function in the rotational dissipation term must satisfy the the moment constraints $c = \langle 1 \rangle$ and $\D = \langle\p\p-\I/2\rangle$. Letting $\mu$ be the largest eigenvalue of $\D/c$, we have $|\D|^2 = 2c^2\mu^2$ so that
\begin{equation}
    \frac{d\cH}{dt} \leq \int_\Omega 2c^2\mu^2 ~ d\x - 4D_T\int_\Omega |\grad c^{1/2}|^2 ~ d\x - 4D_R\int_\Omega c\Big(\int_0^{2\pi}|\partial_\theta\Psi_\mu^{1/2}|^2 ~ d\theta \Big) d\x,
\end{equation}
where at each point in space 
\begin{equation}
    \Psi_\mu = \argmin_{\Psi}\Bigg\{ \int_0^{2\pi}|\partial_\theta\Psi^{1/2}|^2 ~ d\theta : \begin{array}{c}
         \int_0^{2\pi} \Psi ~ d\theta = 1,  \\
          \int_0^{2\pi} \frac{1}{2}\cos2\theta\Psi ~ d\theta = \mu.
    \end{array}\Bigg\}
\end{equation}
Note that we factored the concentration $c$ out of the orientation integral in the last term. Reformulated in terms of $\phi = \Psi^{1/2}$, the minimization problem can be written as
\begin{equation}
    \phi_\mu = \argmin_{\phi}\Bigg\{ \int_0^{2\pi}|\phi'|^2 ~ d\theta : \begin{array}{c}
         \int_0^{2\pi} \phi^2 ~ d\theta = 1,  \\
          \int_0^{2\pi} \frac{1}{2}\cos2\theta\phi^2 ~ d\theta = \mu,
    \end{array}\Bigg\}
\end{equation}
where primes denotes derivatives with respect to $\theta$. The minimizer $\phi_\mu$ satisfies the corresponding Euler-Lagrange equation, which for this case is the Mathieu equation,
\begin{equation}
    \phi_\mu'' + (a - 2q\cos2\theta)\phi_\mu = 0,\label{eq:phi_mu}
\end{equation}
where $a$ and $q$ are Lagrange multipliers that enforce the moment constraints on $\phi_\mu^2$. (A proof of this closely follows that with one constraint in \cite{Evans2010}, Chapter 8.)

\begin{figure}
    \centering
    \includegraphics[scale=0.5]{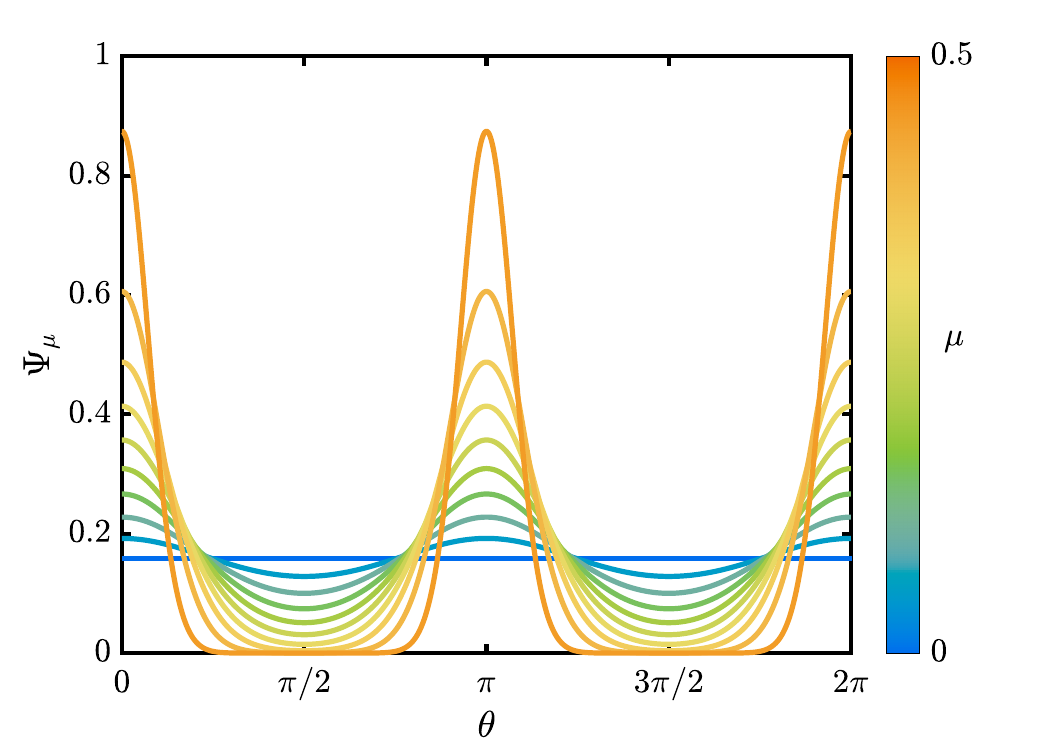}
    \caption{Minimizing distribution $\Psi_\mu$ of rotational dissipation subject to constraints on the largest eigenvalue $\mu$ of $\D/c$. The distribution is bimodal with peaks whose amplitudes increase monotonically with $\mu$.}
    \label{fig:psi_mu}
\end{figure}

The Mathieu equation admits a family of solutions, both periodic and non-periodic, that depend on the values of $a$ and $q$ \citep{Arfken2011}. For each $q$, there is a set of characteristic numbers $a = a_n(q)$ and $a = b_{n+1}(q)$, $n = 0,1,2,\ldots$, such that there are two orthogonal $\pi$-periodic solutions and two orthogonal $2\pi$-periodic solutions, respectively. Numerical computation shows the minimizing function corresponds to the smallest characteristic number $a = a_0(q)$, whose corresponding Mathieu function $\ce_0(\theta;q)$ is $\pi$-periodic, where $q$ is itself a function of $\mu$. The minimizing distribution is then given by $\Psi_\mu =  [\text{ce}_0(\theta;q(\mu))]^2$, and is shown in figure \ref{fig:psi_mu} for several values of $\mu$. Note that as $\mu\rightarrow1/2$, which is the sharply aligned state, the distribution $\Psi_\mu$ approaches a sum of delta functions each located at $\theta = 0$ and $\theta = \pi$.

Because $\phi_\mu$ is $\pi$-periodic so is $\Psi_\mu = \phi_\mu^2$, hence the optimal log-Sobolev constant for functions of this form is $1/2$ (Appendix \ref{app:periodic-sobolev}). We therefore have the bound
\begin{equation}
\int_S \Psi_\mu\log\Big(\frac{\Psi_\mu}{\Psi_0}\Big) ~ d\p \leq \frac{1}{2}\int_S |\grad_p \Psi_\mu^{1/2}|^2 ~ d\p.\label{eq:log-Sobolev-Mathieu}
\end{equation}
Because the model is apolar, if the initial distribution is $\pi$-periodic it will remain $\pi$-periodic and this bound will hold generically. However we need not assume this, and in fact the same distribution will be a minimizer for motile suspensions whose relative entropy also evolves according to Eq. (\ref{eq:dH/dt}) on periodic domains.

The relative entropy of the Mathieu function on the left of (\ref{eq:log-Sobolev-Mathieu}) is challenging to work with as it lacks a clear analytical form. However, a standard variational calculation shows
\begin{equation}
    \frac{1}{c}\int_S \Psi_B \log\Big(\frac{\Psi_B}{c\Psi_0}\Big) ~ d\p \leq \int_S \Psi_\mu\log\Big(\frac{\Psi_\mu}{\Psi_0}\Big) ~ d\p,
\end{equation}
where $\Psi_B = Z^{-1}\exp(\B:\p\p)$ is the maximum entropy, or Bingham, distribution, which minimizes the relative configuration entropy subject to the pointwise constraints $\int_S\Psi ~ d\p = c$ and $\int_S (\p\p-\I/2)\Psi ~ d\p = \D$ \citep{Cover2012}. The parameters $Z(\x,t)$ and $\B(\x,t)$ in the Bingham distribution are Lagrange multipliers that arise from the moment constraints. This distribution function has frequently been applied in closure models of both active and passive suspensions, where it demonstrates good agreement with the kinetic theory in both its linearized behavior as well as long-time nonlinear dynamics \citep{Chaubal1998,Gao2017,Weady2022a,Weady2022b,Freund2023}. Moreover, its analytical properties are well characterized \citep{Li2015}. 

Because the relative entropy is invariant to translations in $\theta$, we may write $\Psi_B = c\exp(\zeta + \xi\cos2\theta)$, where $\zeta(\mu)$ and $\xi(\mu)$ are chosen to satisfy the moment constraints. It is straightforward to show that $\mu = I_1(\xi)/2I_0(\xi)$ and $\zeta = \log(2\pi I_0(\xi))$, where $I_k$ is the modified Bessel function of the first kind \citep{Weady2022a}. In terms of these parameters, we thus have the inequality
\begin{equation}
    \frac{d\cH}{dt} \leq \int_\Omega 2c^2\mu^2 - 4D_T |\grad c^{1/2}|^2 - 8D_Rc\eta ~ d\x,\label{eq:dH/dt_loc}
\end{equation}
where we've introduced the pointwise relative entropy of the Bingham distribution $\eta = (\zeta-\zeta_0) + 2\mu\xi$ with $\zeta_0 = \log(\Psi_0)$. This inequality depends only on the concentration $c$ and the largest eigenvalue $\mu$ of the nematic tensor $\D/c$, and can be treated using elementary methods.

\section{Uniform bounds, nonlinear stability, and time-averaged order parameters}

In this section we use inequality (\ref{eq:dH/dt_loc}) to derive various bounds on the relative entropy and orientational order parameters. We first show the relative entropy is uniformly bounded in time. For sufficiently large rotational diffusion, we show suspensions are nonlinearly stable and lose stability through a supercritical bifurcation. Finally, a similar analysis admits bounds on infinite time averages which hold in the unstable regime.

\subsection{Uniform-in-time bounds on the relative entropy}

Consider the inequality from the previous section
\begin{equation}
    \frac{d\cH}{dt} \leq \int_\Omega 2c^2\mu^2 - 4D_T |\grad c^{1/2}|^2 ~ d\x - 4D_R\int_\Omega \int_S |\grad_p\Psi^{1/2}|^2 ~ d\p d\x,
\end{equation}
where we've retained the rotational dissipation term in its original form. Applying the logarithmic Sobolev inequality (\ref{eq:log-Sobolev}) with the general constant $C = 2$, we have
\begin{equation}
    \frac{d\cH}{dt} \leq \int_\Omega 2c^2\mu^2 + 2D_Rc\log c - 4D_T |\grad c^{1/2}|^2  ~ d\x - 2D_R\cH.
\end{equation}
Thus, if we can show
\begin{equation}
\cF[c,\mu] := \int_\Omega 2c^2\mu^2 + 2D_Rc\log c - 4D_T |\grad c^{1/2}|^2  ~ d\x
\end{equation}
is uniformly bounded, then Gr\"onwall's inequality implies $\cH$ is uniformly bounded as well. Using the trivial inequality $\mu \leq 1/2$, we have
\begin{equation}
\begin{aligned}
\cF[c,\mu] &\leq \int_\Omega \frac{c^2}{2} + 2D_Rc\log c - 4D_T |\grad c^{1/2}|^2 ~ d\x
\\ & \leq 
\Big(\frac{1}{2} + 2D_R\Big)\int_\Omega c^2 ~ d\x - 2D_R|\Omega|,
\end{aligned}
\end{equation}
where we used the inequality $c\log c \leq c(c-1)$ and eliminated the strictly negative spatial dissipation term. It is straightforward to show $\int_\Omega c^2 ~ d\x$ is strictly decreasing, in which case we have the uniform bound
\begin{equation}
\cH(t) \leq \max\Bigg\{\Big[\Big(1 + \frac{1}{4D_R}\Big)\Big(\int_\Omega c_0^2 ~ d\x\Big) - |\Omega|\Big],\cH_0\Bigg\},\label{eq:H_unif}
\end{equation}
where $\cH_0 = \cH(0)$ and $c_0 = c(\x,0)$. For the rest of this section we assume $c_0 = 1$ in which case $c = 1$ for all time.

\subsection{Nonlinear stability of the isotropic state}

From Eq. (\ref{eq:dH/dt_loc}) we have the inequality
\begin{equation}
    \frac{d\cH}{dt} \leq \int_\Omega 2\mu^2 -  8D_R\eta ~ d\x.\label{eq:G_max}
\end{equation}
We claim $4\mu^2 \leq \eta$. To this end, let $h(\mu) = \eta - 4\mu^2$. Differentiating $h$ with respect to $\mu$, applying the identity $d\eta/d\mu = 2\xi$ (Appendix \ref{app:eta}), and using the fact that $\xi \geq 4\mu$ (Inequality (2.21) in \cite{Ifantis1990}), we have
\begin{equation}
    \frac{dh}{d\mu} = 2\xi - 8\mu \geq 0.
\end{equation}
Since $h(0) = 0$, this shows $h$ is strictly non-negative and so the claim holds. Applying this inequality to the integral (\ref{eq:G_max}), we find
\begin{equation}
    \frac{d\cH}{dt} \leq \Big(\frac{1-16D_R}{2}\Big)\int_\Omega \eta ~ d\x.
\end{equation}
%
%
%
Since $\eta \geq 0$, this shows that whenever $D_R > 1/16$ we have $d\cH/dt \leq 0$, with equality only in the globally isotropic state $\mu(\x) = \eta(\mu(\x)) = 0$. This implies $\cH$ is a Lyapunov functional and so the isotropic state is globally attracting, or nonlinearly stable. Remarkably, this condition is independent of the boundary geometry. Linear analysis shows, for sufficiently low translational diffusion, the isotropic state is unstable to infinitesimal perturbations for $D_R < 1/16$ so that this threshold is sharp. Because the nonlinear and linear stability thresholds coincide, this implies the isotropic state loses stability through a supercritical bifurcation.  Moreover, because the dimensionless rotational diffusion coefficient is inversely proportional to the active stress coefficient $|\sigma_a|$, physically this stability threshold can be crossed by increasing activity.

This inequality also provides a bound on the growth rate of $\cH$ when $D_R$ is below the stability threshold. In particular, because $\int_\Omega \eta ~ d\x \leq \cH$, which follows from the fact that the Bingham distribution minimizes the relative entropy, for $D_R < 1/16$ we have $d\cH/dt \leq (1 - 16D_R)\cH/2$. Gr\"onwall's inequality then implies ${\cH(t) \leq \cH_0\exp[(1-16D_R)t/2]}$, however this does not provide an accurate bound on long-time solutions.

\begin{figure}
    \centering
    \includegraphics[scale=0.5]{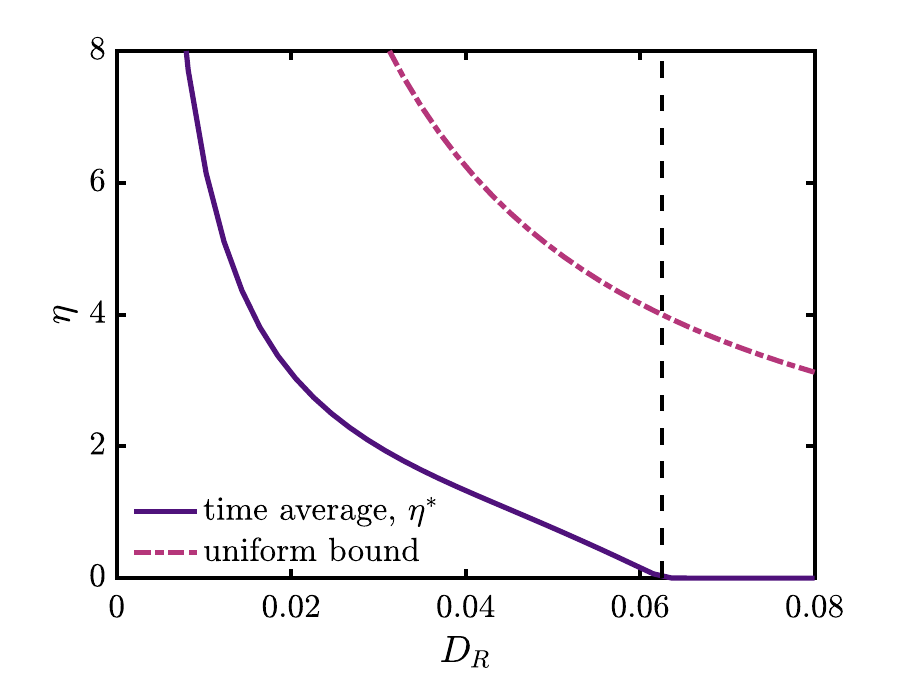}
    \caption{Bounds on the relative entropy of the Bingham distribution. The solid line shows the time-averaged bound $\eta^*(D_R)$ and the dot-dashed line shows the spatially averaged uniform-in-time bound. Beyond the nonlinear stability threshold $D_R > 1/16$ (dashed line), the bound is zero, consistent with nonlinear stability. As $D_R\rightarrow 0$, the bound diverges.}
    \label{fig:fig2}
\end{figure}

\subsection{Bounds on time-averaged order parameters}

In the case $D_R < 1/16$ for which solutions exhibit instabilities, the previous analysis is inconclusive. However, we can exploit the fact the $\cH$ is uniformly bounded in time to derive sharper estimates on long-time behavior. Defining $\overline{\Phi} = \lim_{T\rightarrow\infty} [T^{-1}\int_0^T \Phi(t) ~ dt]$ to be the infinite time average and using the fact that $\cH$ is bounded, taking the infinite time average of both sides of Eq. (\ref{eq:G_max}) gives
\begin{equation}
0 \leq \overline{\int_\Omega 2\mu^2 - 8D_R\eta ~ d\x}.
\end{equation}
Letting $\eps \in (0,1)$ and adding $8D_R\eps (\overline{\int_\Omega \eta ~ d\x})$ to both sides gives 
\begin{equation}
\overline{\int_\Omega \eta ~ d\x} \leq \frac{1}{8D_R\eps}\Bigg[\overline{\int_\Omega 2\mu^2 - 8D_R(1-\eps)\eta ~ d\x}\Bigg].    
\end{equation}
Critical points of the right hand side occur when $\mu = 4D_R(1-\eps)\xi(\mu)$, which is shown by differentiating the integrand with respect to $\mu$. When $D_R < 1/16$, in addition to the trivial solution $\mu = 0$, there is one non-trivial solution to this equation and this solution is the maximum. Therefore, maximizing the right hand side over $\mu$ and minimizing over $\eps$ implies
\begin{equation}
    \overline{\frac{1}{|\Omega|}\int_\Omega \eta ~ d\x}\leq \eta^*(D_R),
\end{equation}
where
\begin{equation}
        \eta^*(D_R) = \inf_{\eps\in(0,1)}\max_{\mu\in[0,1/2]}\Big\{\frac{1}{8D_R\eps}\Big[2\mu^2 - 8D_R(1-\eps) \eta\Big]\Big\}.\label{eq:C(D_R)}
\end{equation}
We solve this minimization problem numerically over $D_R$, with the solution shown in figure \ref{fig:fig2}, including the uniform bound (\ref{eq:H_unif}) for comparison. For $D_R \geq 1/16$, we find $\eta^* = 0$ as required by nonlinear stability. On the other hand, as $D_R\rightarrow 0$ we find $\eta^*\rightarrow\infty$, similar to the uniform bound. This is unavoidable, as any spatially homogeneous distribution is a steady solution of the kinetic theory in this limit and can be taken to have arbitrarily large relative entropy.

The upper bound (\ref{eq:C(D_R)}) is for the relative entropy of the Bingham distribution and does not provide a bound on the relative entropy of the kinetic theory. However, we can use this to derive bounds on the nematic order parameter associated with the true distribution function. Because $\eta$ is a convex function of $\mu$ (Appendix \ref{app:eta}), Jensen's inequality implies
\begin{equation}
\eta\Bigg(\overline{\frac{1}{|\Omega|}\int_\Omega \mu ~ d\x}\Bigg) \leq \overline{\frac{1}{|\Omega|}\int_\Omega \eta ~ d\x} \leq \eta^*(D_R).\label{eq:eta-convex}
\end{equation}
Monotonicity of $\eta$ (Appendix \ref{app:eta}) in turn implies
\begin{equation}
    \overline{\frac{1}{|\Omega|}\int_\Omega \mu ~ d\x} \leq \mu^*(D_R),\label{eq:mu_star}
\end{equation}
where $\mu^*$ solves $\eta(\mu^*(D_R)) = \eta^*(D_R)$, which can again be determined numerically.

\subsection{Numerical verification}

To assess these theoretical bounds, we perform numerical simulations of the kinetic theory in a two-dimensional periodic domain. The numerical method is based on a pseudo-spectral discretization of both the spatial and orientational degrees of freedom of the distribution function, and employs a second-order, implicit-explicit time-stepping scheme. In all simulations we use $512^2$ spatial Fourier modes and $128$ orientational modes, along with the 2/3 anti-aliasing rule. The initial condition is a plane wave perturbation from isotropy $\Psi = 1/2\pi$.

\begin{figure}
    \centering
    \includegraphics[width=\linewidth]{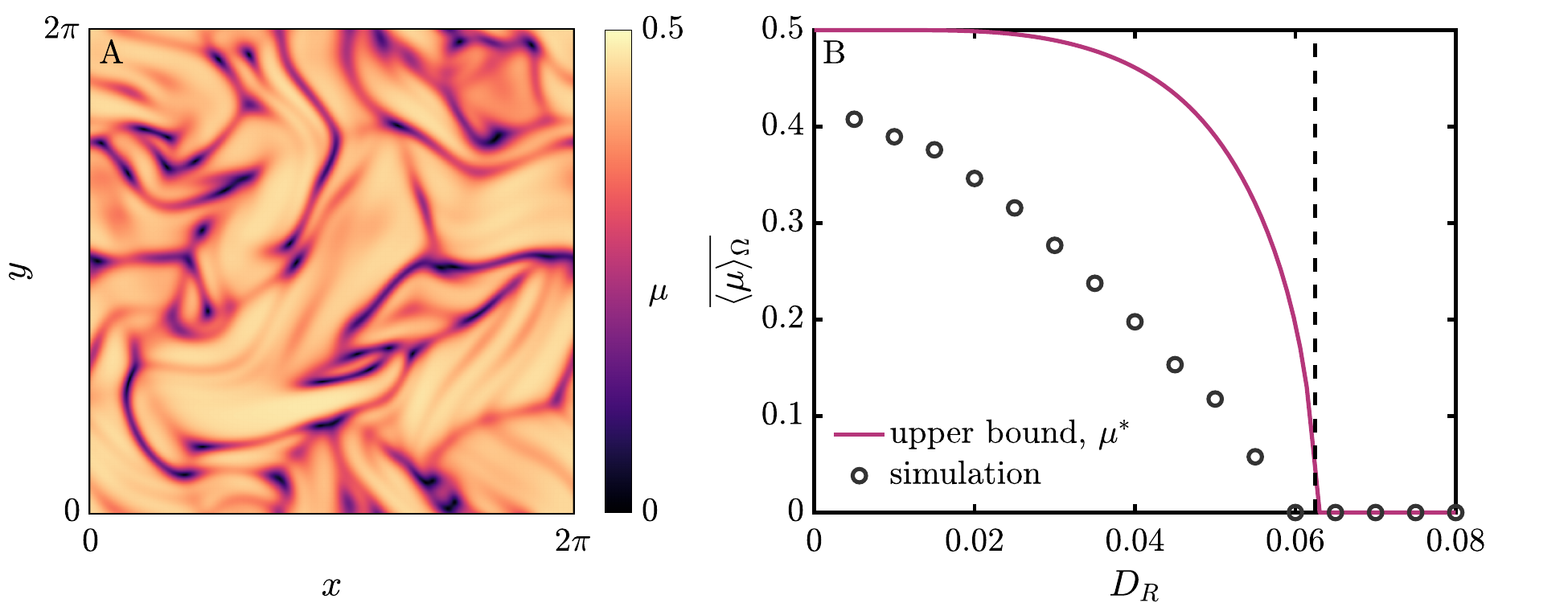}
    \caption{A. Scalar nematic order parameter $\mu$ for a simulation with $D_R = 0.01$ and $D_T = 10^{-3}$. B. Time-averaged value $\overline{\langle\mu\rangle_\Omega} =\overline{(\int_\Omega \mu ~ d\x)/|\Omega|}$ compared to the theoretical bound $\mu^*$ for several values of $D_R$. The bound holds across all $D_R$, and is sharp past the nonlinear stability threshold (dashed line).}
    \label{fig:fig3}
\end{figure}

Figure \ref{fig:fig3}A shows a snapshot of the scalar nematic order parameter $\mu$ for a simulation with $D_R = 0.01$ and $D_T = 10^{-3}$. The nematic order parameter consists of broad regions of high order $\mu\approx0.5$, with narrow defect regions of low order $\mu\approx0$. Figure \ref{fig:fig3}B shows the spatiotemporal average of the nematic order parameter $\overline{\langle\mu\rangle_\Omega} = (1/|\Omega|)\overline{\int_\Omega\mu ~ d\x}$ for simulations with various values of $D_R$ compared to the theoretical bound $\mu^*$ as defined by Eqs. (\ref{eq:eta-convex})-(\ref{eq:mu_star}). The average increases with decreasing $D_R$ in agreement with the upper bound, though the bound does not appear to be sharp in the unsteady regime.

\section{Discussion}

This work provides analytical insight into the nonlinear dynamics of active nematic suspensions. Drawing on analogies with turbulent flows, the entropy method allowed us to derive rigorous bounds on the relative entropy and its fluctuations, establish a criterion for nonlinear stability, and bound time averages of orientational order parameters with relative ease. This analysis showed, under the provided boundary conditions, that the boundary geometry has no influence on stability for immotile suspensions, nor on time-averaged quantities. A key component of our analysis was the use of the Bingham distribution, whose properties made it possible to derive analytically tractable bounds on the rate of dissipation. All of these results extend naturally to three dimensions with slightly modified constants, though the optimal constant in the variational bound on rotational dissipation in Section \ref{sec:rotation} requires more detailed calculation.

The entropy method is flexible and can be applied to other continuum models of active suspensions such as those that involve steric interactions \citep{Ezhilan2013,Gao2017b}, run and tumble dynamics \citep{Subramanian2009}, or chemotaxis \citep{Lushi2012}. In these models other steady states may exist, in which case the relative entropy must be measured as a departure from another steady, possibly inhomogeneous, state $\tilde\Psi = \tilde\Psi(\x,\p)$,
\begin{equation}
    \tilde\cH(t) = \int_\Omega \int_S \Psi \log\Big(\frac{\Psi}{\tilde\Psi}\Big) ~ d\p d\x.
\end{equation}
This form of the relative entropy is also non-negative and vanishes only when $\Psi = \tilde\Psi$, however its rate of change involves additional terms that may include boundary contributions. The evolution of $\tilde\cH$ may also depend on higher order orientational moments of the distribution function, which could modify the constant in the logarithmic Sobolev inequality for minimizers of dissipation.

Motile suspensions in confinement are particularly interesting, since the isotropic state is not a valid solution as it violates the no-flux boundary condition $\partial\Psi/\partial n = (\beta/D_T)(\p\cdot\hat\n)\Psi$, where $\beta$ is the swimming speed \citep{Ezhilan2015}. Moreover, details of the particle geometry may need to be considered to account for admissible configurations near the boundary \citep{Nitsche1990,Schiek1995,Chen2021}. Here the steady state solution will depend on both the boundary geometry and the ratio of motility to spatial diffusion $\beta/D_T$, though the solution lacks a precise analytical form even in simple geometries. Such solutions, as well as their unsteady counterparts, are characterized by concentration boundary layers \citep{Rothschild1963,Berke2008,Elgeti2013}.

We emphasize the results here are limited to immotile suspensions, though numerical and experimental work have found motility has a weak effect on turbulent statistics \citep{Stenhammar2017,Peng2021}. The central challenge posed by particle motility is the lack of a maximum principle on concentration fluctuations from which the estimate (\ref{eq:dH/dt_loc}) can be shown to have no upper bound. Nonetheless, numerical evidence suggests motile suspensions are stable at a threshold near the one derived here, and further work should see if the methods can be extended to determine more general bounds. 

\hspace{0.125in}

\noindent{\bf Acknowledgments}. We thank Gilles Francfort and David Stein for useful discussions, and thank Michael Shelley for encouraging the publication of this work.\\

\noindent{\bf Declaration of interests}. The authors report no conflict of interest.

\appendix

\section{Improved constants for $2\pi/n$-periodic functions}\label{app:periodic-sobolev}

Suppose $f(\theta)$ is $2\pi/n$-periodic, that is $f(\theta+2\pi/n) = f(\theta)$, and define $g = f(\theta/n)$. Assume also that $\int_0^{2\pi} f ~ d\theta = 1$. Making the change of variables $\theta' = \theta/n$, we find
\begin{equation}
\begin{aligned}
    \int_0^{2\pi} g(\theta) ~ d\theta = \int_0^{2\pi} f(\theta/n) ~ d\theta
    = 
    n\int_0^{2\pi/n} f(\theta') ~ d\theta'
    =
    1,
\end{aligned}
\end{equation}
where we used the fact $f(\theta+2\pi/n) = f(\theta)$. Define $\cH[f] = \int_0^{2\pi} f\log(2\pi f) ~ d\theta$ and $\cI[f] = \int_0^{2\pi} |\partial_\theta f^{1/2}|^2 ~ d\theta$. Similar to before, we have
\begin{equation}
\cH[g] = \int_0^{2\pi} g(\theta)\log(2\pi g(\theta)) ~ d\theta = n\int_0^{2\pi/n} f(\theta')\log(2\pi\theta') ~ d\theta' = \cH[f].
\end{equation}
Applying the logarithmic Sobolev inequality to $g$ with the constant $C = 2$ gives
\begin{equation}
    \cH[g] \leq 2\cI[g].
\end{equation}
Evaluating $\cI[g]$, we find
\begin{equation}
\begin{aligned}
    \cI[g] &= \int_0^{2\pi} |\partial_\theta g^{1/2}(\theta)|^2 ~ d\theta
    \\&= \frac{1}{n}\int_0^{2\pi/n} |\partial_{\theta'} f^{1/2}(\theta')|^2 ~ d\theta'
    \\ & =
    \frac{1}{n^2}\cI[f].
\end{aligned}
\end{equation}
We therefore have the inequality
\begin{equation}
    \cH[f] \leq \frac{2}{n^2}\cI[f].
\end{equation}
This inequality is optimal by considering the function $f_n = (1 + \eps \cos n\theta)/2\pi$ as $\eps\rightarrow 0$.

\section{Convexity and monotonicity of $\eta$}\label{app:eta}

Let $\Psi_B = \exp(\zeta + \xi\cos2\theta)$ be the Bingham distribution and consider its relative entropy density $\eta = (\zeta-\zeta_0) + 2\mu\xi$, where $\zeta = -\log(2\pi I_0(\xi))$. Differentiating with respect to $\mu$ gives
\begin{equation}
    \frac{d\eta}{d\mu} = \frac{d\zeta}{d\mu} + 2\xi + 2\mu\frac{d\xi}{d\mu} = 2\xi
\end{equation}
and
\begin{equation}
    \frac{d^2\eta}{d\mu^2} = 2\frac{d\xi}{d\mu},
\end{equation}
where we used $\zeta' = -2\mu\xi'$. Since $\xi\geq 0$, we have $d\eta/d\mu \geq 0$ and so $\eta$ is a monotonic function of $\mu$. Using the identity $\mu = I_1(\xi)/2I_0(\xi)$, we find
\begin{equation}
    2 = \Big[\frac{1}{2} + \frac{1}{2}\frac{I_2(\xi)}{I_0(\xi)} - \Big(\frac{I_1(\xi)}{I_0(\xi)}\Big)^2\Big]\frac{d\xi}{d\mu},\label{eq:dxi/dmu}
\end{equation}
where we've also used $I_1'(\xi) = (I_2(\xi) + I_0(\xi))/2$.
It is sufficient to show the bracketed term in Eq. (\ref{eq:dxi/dmu}) is positive from which we can conclude $d\xi/d\mu > 0$ and hence $d^2\eta/d\mu^2 > 0$, implying $\eta$ is convex. Now define $S = Z^{-1}\int_0^{2\pi} \cos^4\theta e^{\xi\cos2\theta} ~ d\theta$ to be the fourth moment of the Bingham distribution. In terms of $\xi$, this can be written as
\begin{equation}
    S = \frac{1}{8}\Big(3 + 4\frac{I_1(\xi)}{I_0(\xi)} + \frac{I_2(\xi)}{I_0(\xi)}\Big).
\end{equation}
Using this to solve for $I_2(\xi)/I_0(\xi)$, we find
\begin{equation}
\begin{aligned}
\frac{1}{2} + \frac{1}{2}\frac{I_2(\xi)}{I_0(\xi)} - \Big(\frac{I_1(\xi)}{I_0(\xi)}\Big)^2 &= -1 - 4\mu + 4S - 4\mu^2
\\ & = 
4S - 4(\mu + 1/2)^2
\\ & = 
4(S - m^2),
\end{aligned}
\end{equation}
where $m = Z^{-1} \int_0^{2\pi} \cos^2\theta e^{\xi \cos 2\theta}$ is the second moment of the Bingham distribution. The variance inequality ${\langle f\rangle^2 \leq \langle f^2\rangle}$ with $f = \cos^2\theta$ implies $m^2 \leq S$, so the claim holds.

\bibliography{refs}
\bibliographystyle{jfm}

\end{document}